\documentclass[
preprint,
superscriptaddress,
amsmath,
amssymb,
amsfonts,
aip,
showpacs,
]{revtex4-1}

\usepackage{graphicx}
\usepackage{dcolumn}
\usepackage{bm}
\usepackage{easyReview}
\usepackage{soulutf8}

\newcommand {\CIO}{CaIrO$_3$}
\newcommand {\CMO}{CaMnO$_3$}

\newcommand {\RS}{$R_\mathrm{S}$}

\setcitestyle{numbers,square}  

\usepackage{hyperref}

\hypersetup{
setpagesize=false,
 bookmarksnumbered=true,%
 bookmarksopen=true,%
 colorlinks=true,%
 linkcolor=blue,
 citecolor=blue,
 urlcolor=blue
}

\makeatletter
\newcommand{\colorcaption}[2][]{%
	\begingroup%
	\renewcommand{\@caption@fignum@sep}{ (Color online). }%
	\caption[#1]{#2}%
	\endgroup%
}
\makeatother
\begin{document}

\preprint{APS/123-QED}

\title{Electric field control of anomalous Hall effect in {\CIO}/{\CMO} heterostructure}

\author{R. Nishino}%
\author{T. C. Fujita}
\email{fujita@ap.t.u-tokyo.ac.jp}

\affiliation{ 
Department of Applied Physics and Quantum Phase Electronics Center, University of Tokyo, Tokyo 113-8656, Japan
}%

\author{M. Kawasaki}
\affiliation{
Department of Applied Physics and Quantum Phase Electronics Center, University of Tokyo, Tokyo 113-8656, Japan
}%
\affiliation{
RIKEN Center for Emergent Matter Science (CEMS), Wako 351-0198, Japan
}

\date{\today}

\begin{abstract}
We demonstrate an electric field control of anomalous Hall effect emerging in {\CIO}/{\CMO} heterostructures. 
We fabricate both electron-type and hole-type carrier samples by tuning epitaxial strain and then control the carrier density in {\CIO} layer via electric double layer gating technique. 
As the Fermi energy of {\CIO} is tuned close to the Dirac line node, anomalous Hall conductivity is enlarged in both carrier-type samples.
This result reveals that the anomalous Hall effect comes from the intrinsic origin reflecting the Dirac like dispersion in {\CIO}. 
We propose that band splitting induced by the interface ferromagnetism yields several band crossing points near the Dirac line node. 
These points play as a source of the Berry curvature and contribute to the anomalous Hall effect.
\end{abstract}
\maketitle

Oxide heterointerfaces exhibit a variety of exotic physical properties due to complex  interplay between charge, spin and orbital degrees of freedom across the interface~\cite{hwang_emergent_2012,huang_interface_2018,yu_interface_2012}.
In particular, interface ferromagnetism driven by charge transfer is one of the well-known examples.
Ferromagnetism emerges at various interfaces between two non-ferromagnetic compounds; for example, manganites with an antiferromagnetic insulator ground state (CaMnO$_3$, SrMnO$_3$) and paramagnetic conductors (CaRuO$_3$,  SrIrO$_3$, etc.)~\cite{takahashi_interface_2001,nichols_emerging_2016,grutter_electric_2015,freeland_charge_2010}.
When these materials with different chemical potentials are adjacent to each other, electrons are injected into originally empty $e_g$ orbitals of Mn$^{4+}$ to adjust both chemical potentials~\cite{nanda_electron_2007}.
This leads to an intrinsic doping of electrons to manganites and assists double exchange interaction. 
As a result of competition between the double-exchange interaction near the interface and superexchange interaction in the bulk region, the system takes a canted antiferromagnetic state and exhibits weak ferromagnetism.

This interface ferromagnetism in turn gives rise to anomalous Hall effect (AHE), which is experimentally verified in SrIrO$_3$/SrMnO$_3$ superlattices with short period~\cite{nichols_emerging_2016}.
Theoretical calculation predicts that the AHE is intrinsic effect where electrons acquire anomalous Hall velocity induced by the Berry curvature~\cite{bhowal_electronic_2019}.
In this context, it is interesting to modulate the strength of spin-orbit interaction or carrier density (i.e., the position of the Fermi energy) by an external electric field, since the magnitude of the anomalous Hall conductivity (AHC) is linked to these parameters~\cite{bhowal_electric_2019}.
However, an electric field modulation is difficult for SrIrO$_3$/SrMnO$_3$ superlattices because the electric field cannot uniformly modulate every interfaces due to the screening effect, which prohibits us from studying the electric field effect on the AHE emerged at the interface.

In this study, we investigate an electric field effect on AHE in CaIrO$_3$/CaMnO$_3$ heterostructures.
For SrIrO$_3$/SrMnO$_3$, AHE was reported only in superlattice structure so far~\cite{nichols_emerging_2016}.
On the other hand, for CaIrO$_3$/CaMnO$_3$, emergence of AHE was reported even in bilayer structure~\cite{lim_emergent_2020}.
The reported carrier density of CaIrO$_3$ is one or two orders of magnitude smaller than that of SrIrO$_3$; around 10$^{17}$ cm$^{-3}$ and 10$^{19}$ cm$^{-3}$ orders in bulk single crystals~\cite{fujioka_strong-correlation_2019} and epitaxial thin films~\cite{masuko_strain-engineering_2019}, respectively, which is more suitable for an electric field control of the Fermi energy. 
Furthermore, {\CIO} is known to be a topological semimetal, so-called nodal line semimetal, and to possess Dirac line node near the Fermi energy, wherein the conduction and valence bands cross along a closed line in momentum space.
This Dirac line node imparts high-mobility carriers due to the large band dispersion as in the case of other Dirac electron systems.
Moreover, by breaking the time-reversal symmetry, degeneracy of the line node can be lifted, yielding Weyl nodes which function as sources of the Berry curvature and contribute to AHE.
These features render this system an ideal platform to examine the Fermi energy dependence of AHE originating from Dirac-like band structure in oxides.
We fabricate heterostructures with \textit{n}- and \textit{p}-type carriers by controlling epitaxial strain imposed by substrates and then modulate the carrier density in {\CIO} layer via electric double layer (EDL) gating method~\cite{nishino_electrical_2018,nishino_ferroelectric_2018,nishino_evolution_2020,ueno_electric-field-induced_2008}.

{\CIO}/{\CMO} heterostructures were grown on (001) oriented LaAlO$_3$ and SrTiO$_3$ substrates by pulsed laser deposition.
CaMnO$_3$ films were grown at 600$\ {}^\circ\mathrm{C}$ under 8 mTorr oxygen pressure.
CaIrO$_3$ films were then deposited \textit{in situ} on the CaMnO$_3$ at 600$\ {}^\circ\mathrm{C}$ under 40 mTorr oxygen pressure.
Typical thickness of CaIrO$_3$ (CaMnO$_3$) for heterostructures is $\sim$6 nm ($\sim$1.5 nm).
Reciprocal space mappings indicate that CaIrO$_3$ films are fully strained to both LaAlO$_3$ and SrTiO$_3$ substrates (see Supplementary Fig. S1).
For comparison of the transport properties between heterostructures and each constituent film, we also prepared 7 nm-thick CaIrO$_3$ and 1.5 nm-thick Ce$_{0.05}$Ca$_{0.95}$MnO$_3$ films on LaAlO$_3$ substrates.
Since electrons may be injected into the CaMnO$_3$ in the heterostructure, we tried to mimic a similar electronic state by Ce substitution.
This 5\% Ce-doping corresponds to the interface charge transfer of 0.1 electron/Mn, which is comparable with those of CaRuO$_3$/CaMnO$_3$ ($\sim$0.08 electron/Mn) in Ref.~\cite{nanda_electron_2007} and CaIrO$_3$/CaMnO$_3$ ($\sim$0.11 electron/Mn) in Ref.~\cite{lim_emergent_2020}.
For the Ce$_{0.05}$Ca$_{0.95}$MnO$_3$ film, 4 nm-thick CaTiO$_3$ epitaxial cap layer was deposited to avoid surface degradation.
For EDL gating experiments, we employed ionic liquid (IL), N,N-diethyl-N-methyl-N-(2-methoxyethyl) N-methylammonium tetrafluoroborate (DEME-BF$_4$), in which a gate electrode of Pt coil was immersed.
The schematic of the device structure is depicted in Fig. S2(a) of Supplementary Materials.
Typical channel size for electrical transport measurements is $2\times4$ $ \mathrm{mm}^2$, with which the measured resistance was converted into sheet resistance ({\RS}).
The transport properties were measured under vacuum with a back pressure of $1\times$10$^{-5}$ Torr. 
Before the electrical measurements, the IL was stored in a vacuum hot plate at 90$\ {}^\circ\mathrm{C}$ for several hours to remove water contamination which may induce some electrochemical reactions.

\begin{figure}[h]
	\includegraphics[width=15cm]{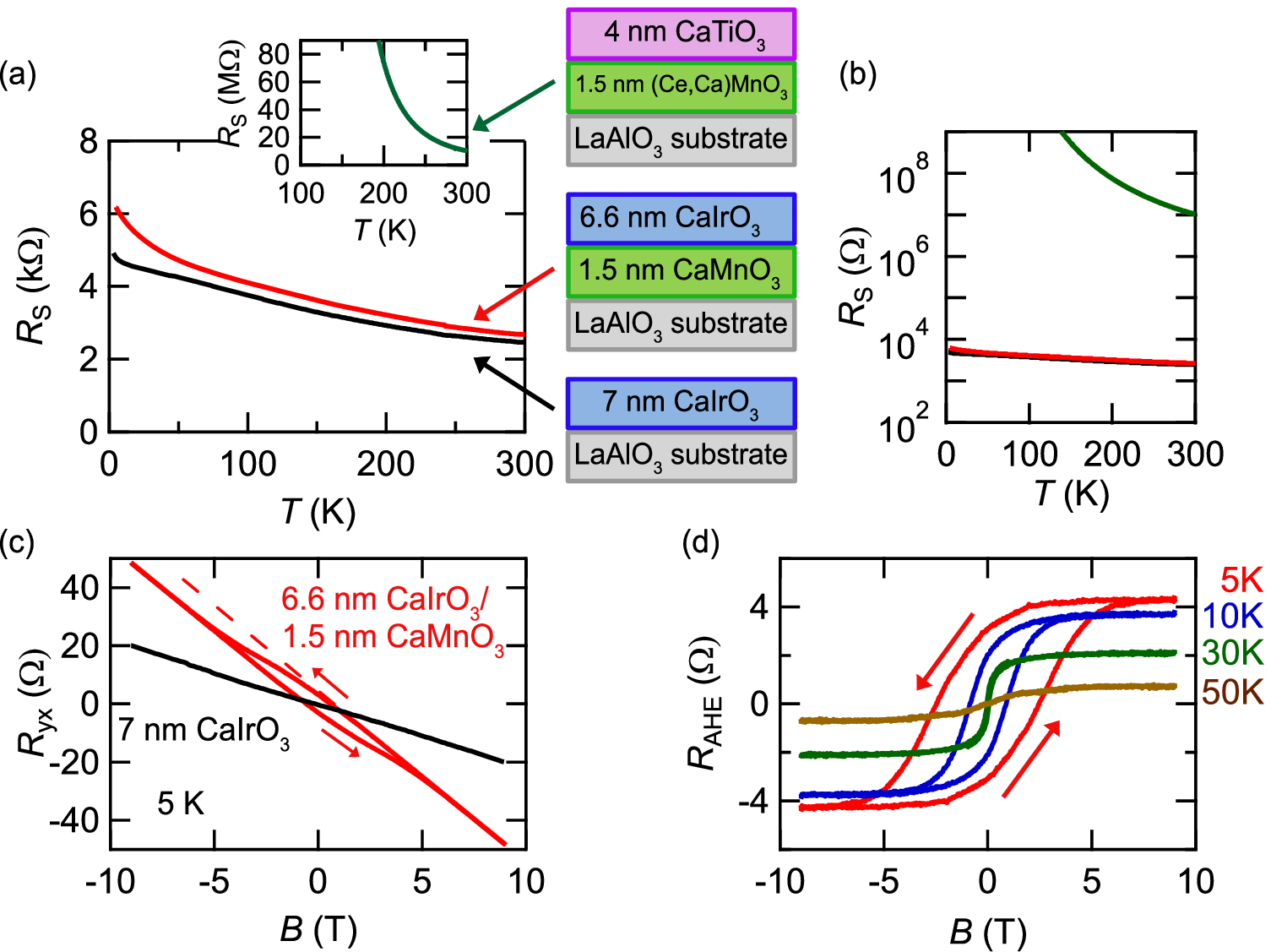}
	\colorcaption{\label{comp}
	Temperature dependence of sheet resistance ($R_\mathrm{S}$) for CaIrO$_3$/CaMnO$_3$ heterostructure (red) and CaIrO$_3$ thin film (black) grown on LaAlO$_3$ substrate.
	The inset shows $R_\mathrm{S}$ for 1.5 nm-thick Ce$_{0.05}$Ca$_{0.95}$MnO$_3$ thin film (green) grown on LaAlO$_3$ substrate as a function of temperature.
	$R_\mathrm{S}$ is displayed in (a) linear and (b) logarithmic scales.
	Schematics are shown to indicate the structure of each sample as well as the thickness of each layer.
	(c) Magnetic field (\textit{B}) dependence of Hall resistance ($R_\mathrm{yx}$) for CaIrO$_3$/CaMnO$_3$ heterostructure (red) and CaIrO$_3$ film (black) at 5 K. 
	The red dashed line represents the linear fitting from the higher magnetic field region. (d) Anomalous Hall resistance ($R_\mathrm{AHE}$) of the heterostructure at various temperatures.
	}
\end{figure}

We first discuss the difference of the transport properties between the heterostructure and each constituent film. 
There has been a report about metallic temperature dependence of resistivity and emerging AHE for compressively strained 20 nm-thick Ce$_{0.05}$Ca$_{0.95}$MnO$_3$ films grown on (001) oriented YAlO$_3$ substrates~\cite{vistoli_giant_2019}.
Therefore, it is important to clarify which layer contributes to the transport phenomena in our heterostructures.
Figure~\ref{comp}(a) shows the temperature dependence of $R_\mathrm{S}$ for the CaIrO$_3$/CaMnO$_3$ heterostructure and the CaIrO$_3$ film grown on LaAlO$_3$ substrates.
The inset shows the $R_\mathrm{S}$ of 1.5 nm-thick Ce$_{0.05}$Ca$_{0.95}$MnO$_3$ film as a function of temperature.
The behavior of $R_\mathrm{S}$ in the heterostructure is similar to that in the CaIrO$_3$ film. 
Both samples exhibit a semimetallic temperature dependence and the $R_\mathrm{S}$ moderately increases with decreasing temperature.
On the other hand, the $R_\mathrm{S}$ of Ce$_{0.05}$Ca$_{0.95}$MnO$_3$ film is several orders of magnitude higher than that of the heterostructure and exhibits an insulating behavior as shown in Fig.~\ref{comp}(b), which is totally different from the Ce$_{0.05}$Ca$_{0.95}$MnO$_3$ films grown on YAlO$_3$ (Ref.\cite{vistoli_giant_2019}).
This contrast plausibly comes from tensile strain imposed on our Ce$_{0.05}$Ca$_{0.95}$MnO$_3$ films grown on LaAlO$_3$, which stabilizes an insulating ground state as reported previously~\cite{xiang_phase_2012}.

In opposition to the similarity of the behavior in $R_\mathrm{S}$ between the heterostructure and the CaIrO$_3$ film, a clear difference is observed in Hall measurements.
Figure~\ref{comp}(c) shows magnetic field (\textit{B}) dependence of Hall resistance ($R_\mathrm{yx}$) at 5 K. While the $R_\mathrm{yx}$ for the CaIrO$_3$ film is merely linear to \textit{B}, the $R_\mathrm{yx}$ for the heterostructure exhibits a clear hysteresis, indicating the emergence of AHE due to the interface ferromagnetism.
In this case, the $R_\mathrm{yx}$ can be empirically expressed by $R_\mathrm{yx}$ = $R_\mathrm{H}B$ + $R_\mathrm{AHE}$, where the $R_\mathrm{H}B$ ($R_\mathrm{H}$: Hall coefficient) and $R_\mathrm{AHE}$ terms denote ordinary and anomalous Hall resistances, respectively.
Figure~\ref{comp}(d) shows $R_\mathrm{AHE}$ as a function of \textit{B} at various temperatures.
The $R_\mathrm{AHE}$ term is extracted by subtracting the $R_\mathrm{H}B$ from the measured $R_\mathrm{yx}$, where the $R_\mathrm{H}B$ term is estimated from the linear fitting in the higher magnetic field region as shown in the red broken line in Fig.~\ref{comp}(c).
$R_\mathrm{AHE}$ emerges below $\sim$60 K and exhibits an anticlockwise hysteresis at low temperature.
It is worth noting that the sign of AHE is positive for the heterostructure while previously reported Ce$_{0.05}$Ca$_{0.95}$MnO$_3$ films on YAlO$_3$ substrates exhibits the negative sign of AHE~\cite{vistoli_giant_2019}.

From the comparison of the transport properties between the heterostructure and each constituent film, we have confirmed that CaIrO$_3$ layer is dominant in both electrical conduction and the observed AHE rather than electron doped CaMnO$_3$ layer in the heterostructure.
Here we propose a possible mechanism of the observed AHE as follows: (i) electrons transfer from CaIrO$_3$ to CaMnO$_3$ layers and induce double-exchange interaction, resulting in a weak ferromagnetism (canted antiferromagnetism) in CaMnO$_3$ layer near the interface, (ii) magnetization is expected to be induced in CaIrO$_3$ near the interface due to magnetic proximity effect, and (iii) CaIrO$_3$ layer with magnetization exhibits the AHE.
In this study, it is difficult to reveal the magnetic ordering at the interface in detail only from the transport properties.
However, recent experimental result and theoretical calculations for SrIrO$_3$/SrMnO$_3$ superlattice revealed that the intra-layer interaction within the SrMnO$_3$ and SrIrO$_3$ layers is ferromagnetic but the inter-layer interaction between the SrMnO$_3$ and SrIrO$_3$ layers is antiferromagnetic~\cite{nichols_emerging_2016,bhowal_electronic_2019,okamoto_charge_2017}.
Although it is intriguing to elucidate the mechanism of the emergent ferromagnetism in CaIrO$_3$/CaMnO$_3$ heterostructures, this is beyond the scope of this report and remains as future work.

\begin{figure}[h]
	\includegraphics[width=6.8cm]{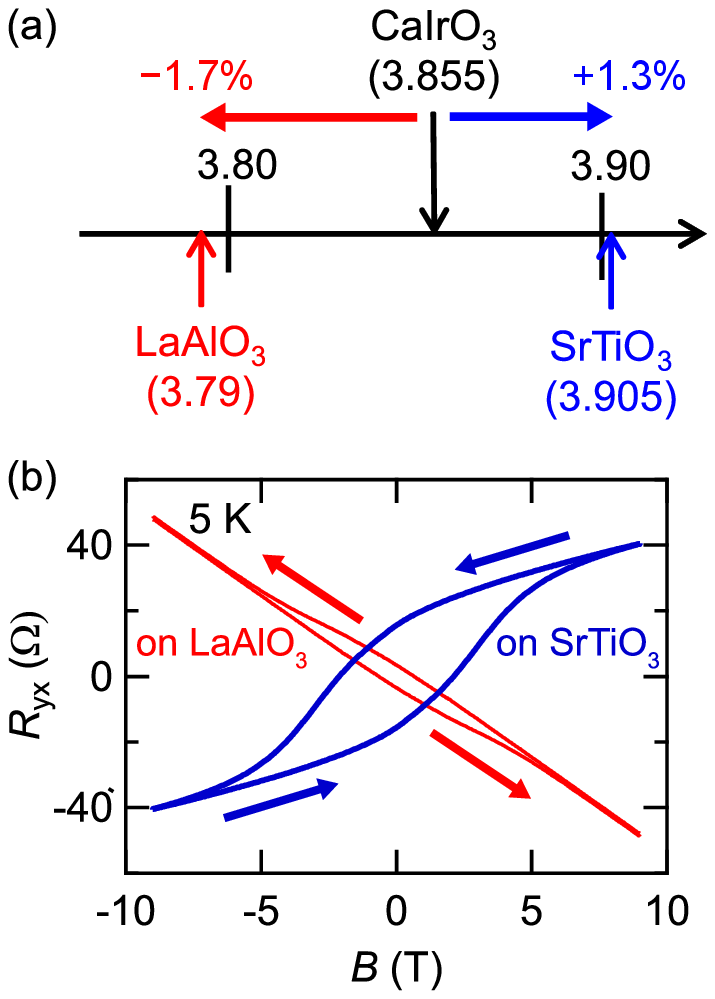}
	\colorcaption {\label{strain}
	(a) Lattice parameters of CaIrO$_3$, LaAlO$_3$ and SrTiO$_3$ substrates.
	Those for CaIrO$_3$ and LaAlO$_3$ are displayed in a pseudo-cubic setting.
	The lattice mismatches between CaIrO$_3$ and the substrates are also indicated.
	(b) Magnetic field (\textit{B}) dependence of Hall resistance ($R_\mathrm{yx}$) at 5 K for CaIrO$_3$/CaMnO$_3$ heterostructures grown on LaAlO$_3$ (red) and SrTiO$_3$ (blue) substrates. 
	}
\end{figure}

We then examine the effect of epitaxial strain on the transport properties in the heterostructures.
As shown in Fig.~\ref{strain}(a), CaIrO$_3$ grown on SrTiO$_3$ substrates is imposed on tensile strain while that on LaAlO$_3$ substrates is compressively strained.
Figure~\ref{strain}(b) shows \textit{B} dependence of $R_\mathrm{yx}$ for the heterostructures grown on SrTiO$_3$ (blue) and LaAlO$_3$ (red) substrates measured at 5 K.
Hall coefficient $R_\mathrm{H}$ exhibits opposite sign between two samples, indicating that hole (electron) type carrier is dominant for the heterostructure grown on SrTiO$_3$ (LaAlO$_3$).
Previous studies report that carrier type of CaIrO$_3$ thin films is sensitive to epitaxial strain ~\cite{masuko_strain-engineering_2019,hirai_semimetallic_2015}.
It has been theoretically predicted that tetragonal distortion can lift the degeneracy of $t_{\mathrm{2g}}$ orbitals in the $J_{\mathrm{eff}}=1/2$ state of Ir$^{4+}$ near Fermi level~\cite{jackeli_mott_2009}, and thus the epitaxial strain might induce this carrier type change.
Yet, considering the sensitivity of the band structure of CaIrO$_3$ against electron correlation as well~\cite{fujioka_strong-correlation_2019}, further elucidation of the origin of the change in carrier type is a matter of speculation.  

\begin{figure*}[h]
	\includegraphics[width=7cm]{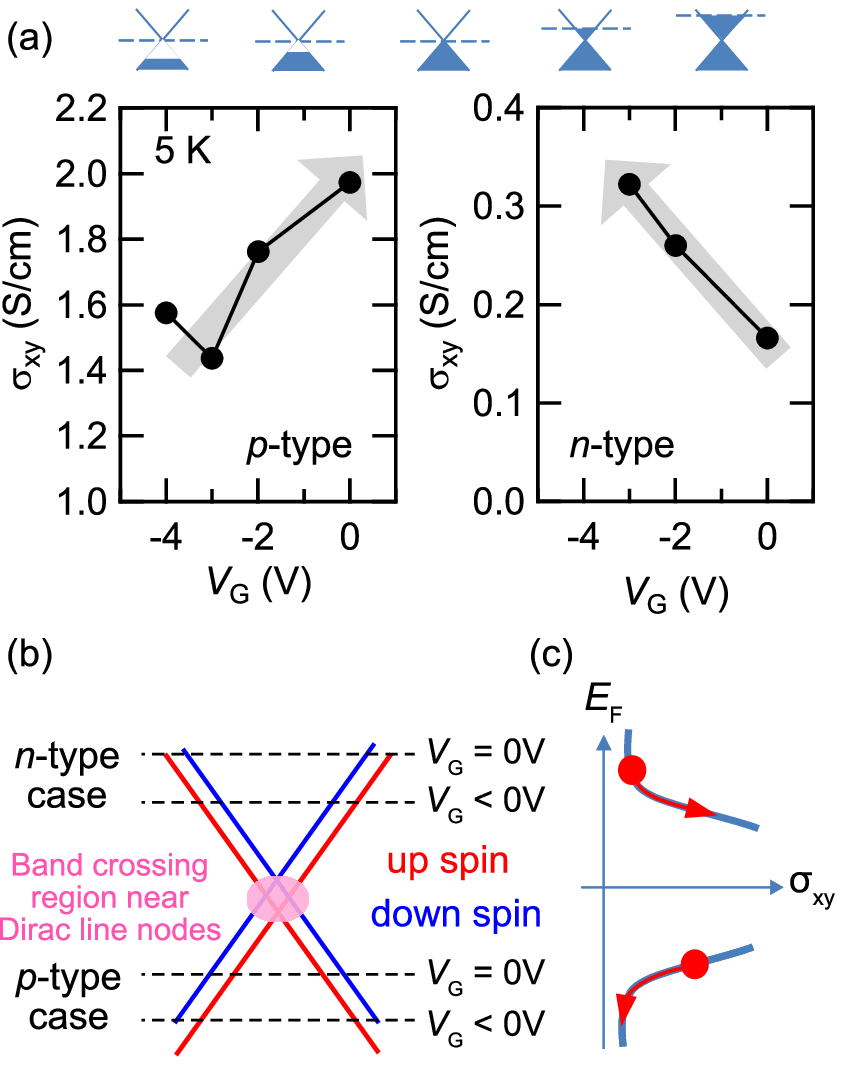}
	\colorcaption{\label{Dirac} 
		(a) Anomalous Hall conductivity ($\sigma_\mathrm{xy}$) at 5 K as a function of gate voltage ($V_\mathrm{G}$).
		The top schematics represents a position of the Fermi energy related to Dirac line node of CaIrO$_3 $.
		The gray arrows represent a guide to the eye.
		(b) Illustration for the band structure of CaIrO$_3$ after band splitting by exchange interaction.
		Red and blue lines represent up and down spin bands, respectively.
		(c) Illustration of the relationship between $\sigma_\mathrm{xy}$ and position of the Fermi energy, where the crossing point of horizontal and vertical axes indicate the band crossing region in (b).
			}
\end{figure*}

To obtain further insight into the origin of the observed AHE, we attempt to tune the position of the Fermi energy of CaIrO$_3$ layer via EDL gating method, where negative gate voltage is applied to the both carrier-type samples.
Negative gate voltage corresponds to tuning the Fermi energy closer to (away from) Dirac line node of CaIrO$_3$ for \textit{n}-type (\textit{p}-type) sample as shown in the top schematics of Fig.~\ref{Dirac}(a).
We performed Hall measurements at several gate voltages.
Each gate voltage was applied at 265 K for 60 mins before the samples were cooled down to each measurement temperature at the rate of 0.5 K/min.
Before Hall measurements, we confirmed that negative gate voltage reversibly modulated $R_\mathrm{S}$ for CaIrO$_3$ and ruled out the possibility of electrochemical reactions (see supplementary Fig. S2(b)).

Figures~\ref{gate}(a) and~\ref{gate}(b) show \textit{B} dependence of $R_\mathrm{AHE}$ at 5 K as a function of gate voltage ($V_\mathrm{G}$) for \textit{p}-  and \textit{n}-type heterostructures, respectively.
The temperature dependence of the $R_\mathrm{S}$ is summarized in supplementary Fig. S3. 
For the \textit{n}-type heterostructure, negative $V_\mathrm{G}$ increases $R_\mathrm{AHE}$.
On the other hand, $R_\mathrm{AHE}$ decreases with negative $V_\mathrm{G}$ for the \textit{p}-type heterostructure.
Figures~\ref{gate}(c) and~\ref{gate}(d) show temperature dependence of $\sigma_\mathrm{xy}$  at 9 T under several $V_\mathrm{G}$ for the \textit{p}- and \textit{n}-type heterostructures, respectively.
Here, $\sigma_\mathrm{xy}$ is calculated as
\begin{equation}\label{sigma}
\sigma_\mathrm{xy} = \frac{R_\mathrm{AHE}t}{(R_\mathrm{S}t)^2+(R_\mathrm{AHE}t)^2}
\end{equation}
where \textit{t} is the thickness of CaIrO$_3$ layer.
In both heterostructures, the ferromagnetic transition temperature, where the AHE emerges, is unchanged ($\sim$60 K).
Furthermore, the coercive field, which is estimated from the hysteresis loop of $R_\mathrm{AHE}$, also remains nearly unchanged against the amplitude of the gate voltage.
These results suggest that the magnetic properties of the heterostructures are not modulated by the EDL gating.
Rather, the electric field only modifies the carrier density of CaIrO$_3$ (i.e., the position of the Fermi energy).
It should be pointed out that the $\sigma_\mathrm{xy}$ calculated from Eq.~(\ref{sigma}) may be underestimated because the effective thickness accounting for the AHE may be smaller than $t\approx6$ nm, if we consider the interfacial ferromagnetism arising from proximity effect.

Next, we discuss how CaIrO$_3$ layer contributes to the AHE.
It is well known that AHE is generally classified into two types~\cite{onoda_quantum_2008}.
One comes from an intrinsic origin where electrons acquire anomalous Hall velocity induced by the Berry curvature in momentum space.
This mechanism is dominant in the moderately dirty system where longitudinal conductivity $\sigma_\mathrm{xx}$ is below $\sim$10$^4$ S/cm.
The other comes from an extrinsic origin where electrons are scattered by magnetic impurities via spin-orbit interaction and contribute to the AHE.
This mechanism is dominant for larger $\sigma_\mathrm{xx}$ above $\sim$10$^5$ S/cm.
Taking into account that $\sigma_\mathrm{xx}$ for our heterostructures is below $\sim$10$^3$ S/cm, we can assume that the AHE of the heterostructures comes from the intrinsic origin.
At this point, the Dirac-like band dispersion of CaIrO$_3$ has significance as it can be a source of the Berry curvature.
In Kubo formula~\cite{nagaosa_anomalous_2010}, anomalous Hall conductivity $\sigma_\mathrm{xy}$ is given by
\begin{equation}\label{kubo}
\begin{split}	
\sigma_\mathrm{xy}&=e^2\hbar\sum_{n\neq n'} \int \frac{d\textbf{\textit{k}}}{(2\pi)^3}[f(\varepsilon_{n}(\textbf{\textit{k}}))-f(\varepsilon_{n'}(\textbf{\textit{k}}))] \\
&  \times \mathrm{Im}  \frac{  \langle n,\textbf{\textit{k}}| v_x(\textbf{\textit{k}}) | n',\textbf{\textit{k}} \rangle \langle n',\textbf{\textit{k}}| v_y (\textbf{\textit{k}})| n,\textbf{\textit{k}} \rangle}{[\varepsilon_{n}(\textbf{\textit{k}}) - \varepsilon_{n'}(\textbf{\textit{k}})]^2}
\end{split}
\end{equation}
where \textit{n} is band index, $f(\varepsilon_{n}(\textbf{\textit{k}}))$ is Fermi distribution function and $\textbf{\textit{v}}(\textbf{\textit{k}})$ is velocity operator defined in the $\textit{\textbf{\textit{k}}}$-dependent Hamiltonian (\textit{H}(\textit{\textbf{\textit{k}}})) for the
periodic part of the Bloch functions by 
\begin{equation}\label{velocity}
 \textbf{\textit{v}}(\textbf{\textit{k}})=\frac{1}{i\hbar}[\textbf{\textit{k}},H(\textit{\textbf{\textit{k}}})]=\frac{1}{\hbar}\nabla_kH(\textit{\textbf{\textit{k}}}).
\end{equation}
Equation~(\ref{kubo}) can be transformed into
\begin{equation}\label{Berry}
\sigma_\mathrm{xy}=-\frac{e^2}{\hbar}\sum_{n\neq n'} \int \frac{d\textbf{\textit{k}}}{(2\pi)^3}f(\varepsilon_{n}(\textbf{\textit{k}}))b^z_{n}(\textbf{\textit{k}})
\end{equation}
where $b^z_{n}(\textbf{\textit{k}})$ is the Berry curvature.
Equation~(\ref{Berry}) indicates that $\sigma_\mathrm{xy}$ is the sum of the Berry curvature over up to the Fermi energy.
According to Eq.~(\ref{kubo}), the anomalous Hall conductivity is enhanced in the following conditions: (i) large group velocity $\textbf{\textit{v}}(\textbf{\textit{k}})$, which is satisfied in a large band dispersion at the Fermi energy and (ii) two bands are energetically close to each other (i.e., small $\varepsilon_{n}(\textbf{\textit{k}}) - \varepsilon_{n'}(\textbf{\textit{k}})$).
These two conditions are indeed satisfied when the Fermi energy of CaIrO$_3$ is tuned close to the Dirac line node. 

Figure~\ref{Dirac}(a) shows gate voltage dependence of $\sigma_\mathrm{xy}$ at 5 K.
The schematics are also indicated for the Fermi energy relative to the Dirac-like dispersion of CaIrO$_3$. 
Both \textit{p}- and \textit{n}-type samples exhibit an almost monotonic increase in $\sigma_\mathrm{xy}$ as the Fermi energy approaches the Dirac line node.
This trend is consistent with Eq.~(\ref{kubo}) and our assumption: the AHE in the heterostructures is the intrinsic effect and Dirac-like dispersion of CaIrO$_3$ contributes to the AHE.
Figure~\ref{Dirac}(b) illustrates our interpretation of the relationship between the band structure of CaIrO$_3$ and its AHE in the heterostructures.
We assume that the degenerated Dirac line node of the CaIrO$_3$ layer is split into up and down spin bands by the exchange interaction which is induced by the magnetic proximity effect from the CaMnO$_3$ layer.
Although the magnitude of exchange energy is uncertain, it may be several meV order since the AHE commonly emerges at $\sim$60 K, or $k_\mathrm{B}T_\mathrm{C}\approx$ 5 meV.
Such an induced band splitting results in the band crossings near Dirac line node and the creation of several Weyl nodes as shown in Fig.~\ref{Dirac}(b).
In this study, we assume that the Fermi energy of CaIrO$_3$ is far above (below) Dirac line node for \textit{n}(\textit{p})-type samples.
It is reported that the Fermi energy for bulk single crystals is about 10 meV above Dirac line node~\cite{fujioka_strong-correlation_2019}.
Since the carrier density of CaIrO$_3$ films is two orders of magnitude higher than that of single crystals~\cite{masuko_strain-engineering_2019}, the Fermi energy of our CaIrO$_3$ layers is estimated to be away from Dirac line node in both carrier-type samples.
In this assumption, the negative $V_\mathrm{G}$ means that Fermi energy approaches (leaves) the band crossings for \textit{n}(\textit{p})-type.
As the Fermi energy is closer to the band crossings, each carrier acquires larger anomalous velocity from the Berry curvatures, resulting in the enhancement of the AHE (Fig.~\ref{Dirac}(c)).

In conclusion, we fabricate CaIrO$_3$/CaMnO$_3$ heterostructures with both-carrier types and confirm that CaIrO$_3$ takes the role of the carrier transport in this system.
We perform an electric field control of the AHE emerging at the heterostructures by using EDL gating.
In both carrier-type heterostructures, anomalous Hall conductivity is enlarged as a gate voltage tunes the Fermi energy closer to Dirac line node of {\CIO}.
This result indicates that the AHE comes from an intrinsic origin reflecting the Dirac-like linear energy dispersion of {\CIO}. 
We propose a plausible explanation for the AHE in the context of the Berry curvature originating from Weyl nodes which are presumably induced by the magnetic proximity effect from the CaMnO$_3$ layer.
Our work provides important insight into the origin and manipulation of AHE in the oxide heterointerfaces with Dirac-like band dispersion.
Also recently, emergence of Dirac electrons has been reported in strained SrNbO$_3$ thin films~\cite{mohanta_semi-dirac_2021,ok_correlated_2021}, where enhancement of AHE is expected under broken time-reversal and inversion symmetries. 
In this sense, the demonstrated technique in this report would be a promising way to modulate AHE by not only tuning Fermi level but also breaking inversion symmetry at the interface.

\section*{Supplementary Material}
See supplementary material for the additional XRD and transport measurements data.

\begin{acknowledgements}
This work was partly supported by the Japan Science and Technology Agency Core Research for Evolutional Science Technology (JST CREST) (No. JPMJCR16F1) and Izumi Science and Technology Foundation.
\end{acknowledgements}

\section*{Data Availability Statement}
The data that support the findings of this study are available from the corresponding author upon reasonable request.

\bibliography{CIO-CMO_NoNote}

\begin{thebibliography}{26}%
\makeatletter
\providecommand \@ifxundefined [1]{%
 \@ifx{#1\undefined}
}%
\providecommand \@ifnum [1]{%
 \ifnum #1\expandafter \@firstoftwo
 \else \expandafter \@secondoftwo
 \fi
}%
\providecommand \@ifx [1]{%
 \ifx #1\expandafter \@firstoftwo
 \else \expandafter \@secondoftwo
 \fi
}%
\providecommand \natexlab [1]{#1}%
\providecommand \enquote  [1]{``#1''}%
\providecommand \bibnamefont  [1]{#1}%
\providecommand \bibfnamefont [1]{#1}%
\providecommand \citenamefont [1]{#1}%
\providecommand \href@noop [0]{\@secondoftwo}%
\providecommand \href [0]{\begingroup \@sanitize@url \@href}%
\providecommand \@href[1]{\@@startlink{#1}\@@href}%
\providecommand \@@href[1]{\endgroup#1\@@endlink}%
\providecommand \@sanitize@url [0]{\catcode `\\12\catcode `\$12\catcode
  `\&12\catcode `\#12\catcode `\^12\catcode `\_12\catcode `\%12\relax}%
\providecommand \@@startlink[1]{}%
\providecommand \@@endlink[0]{}%
\providecommand \url  [0]{\begingroup\@sanitize@url \@url }%
\providecommand \@url [1]{\endgroup\@href {#1}{\urlprefix }}%
\providecommand \urlprefix  [0]{URL }%
\providecommand \Eprint [0]{\href }%
\providecommand \doibase [0]{http://dx.doi.org/}%
\providecommand \selectlanguage [0]{\@gobble}%
\providecommand \bibinfo  [0]{\@secondoftwo}%
\providecommand \bibfield  [0]{\@secondoftwo}%
\providecommand \translation [1]{[#1]}%
\providecommand \BibitemOpen [0]{}%
\providecommand \bibitemStop [0]{}%
\providecommand \bibitemNoStop [0]{.\EOS\space}%
\providecommand \EOS [0]{\spacefactor3000\relax}%
\providecommand \BibitemShut  [1]{\csname bibitem#1\endcsname}%
\let\auto@bib@innerbib\@empty
\bibitem [{\citenamefont {Hwang}\ \emph {et~al.}(2012)\citenamefont {Hwang},
  \citenamefont {Iwasa}, \citenamefont {Kawasaki}, \citenamefont {Keimer},
  \citenamefont {Nagaosa},\ and\ \citenamefont {Tokura}}]{hwang_emergent_2012}%
  \BibitemOpen
  \bibfield  {author} {\bibinfo {author} {\bibfnamefont {H.~Y.}\ \bibnamefont
  {Hwang}}, \bibinfo {author} {\bibfnamefont {Y.}~\bibnamefont {Iwasa}},
  \bibinfo {author} {\bibfnamefont {M.}~\bibnamefont {Kawasaki}}, \bibinfo
  {author} {\bibfnamefont {B.}~\bibnamefont {Keimer}}, \bibinfo {author}
  {\bibfnamefont {N.}~\bibnamefont {Nagaosa}}, \ and\ \bibinfo {author}
  {\bibfnamefont {Y.}~\bibnamefont {Tokura}},\ }\href {\doibase
  10.1038/nmat3223} {\bibfield  {journal} {\bibinfo  {journal} {Nature
  Materials}\ }\textbf {\bibinfo {volume} {11}},\ \bibinfo {pages} {103}
  (\bibinfo {year} {2012})}\BibitemShut {NoStop}%
\bibitem [{\citenamefont {Huang}\ \emph {et~al.}(2018)\citenamefont {Huang},
  \citenamefont {{Ariando}}, \citenamefont {Renshaw~Wang}, \citenamefont
  {Rusydi}, \citenamefont {Chen}, \citenamefont {Yang},\ and\ \citenamefont
  {Venkatesan}}]{huang_interface_2018}%
  \BibitemOpen
  \bibfield  {author} {\bibinfo {author} {\bibfnamefont {Z.}~\bibnamefont
  {Huang}}, \bibinfo {author} {\bibnamefont {{Ariando}}}, \bibinfo {author}
  {\bibfnamefont {X.}~\bibnamefont {Renshaw~Wang}}, \bibinfo {author}
  {\bibfnamefont {A.}~\bibnamefont {Rusydi}}, \bibinfo {author} {\bibfnamefont
  {J.}~\bibnamefont {Chen}}, \bibinfo {author} {\bibfnamefont {H.}~\bibnamefont
  {Yang}}, \ and\ \bibinfo {author} {\bibfnamefont {T.}~\bibnamefont
  {Venkatesan}},\ }\href {\doibase 10.1002/adma.201802439} {\bibfield
  {journal} {\bibinfo  {journal} {Advanced Materials}\ }\textbf {\bibinfo
  {volume} {30}},\ \bibinfo {pages} {1802439} (\bibinfo {year}
  {2018})}\BibitemShut {NoStop}%
\bibitem [{\citenamefont {Yu}\ \emph {et~al.}(2012)\citenamefont {Yu},
  \citenamefont {Luo}, \citenamefont {Yi}, \citenamefont {Zhang}, \citenamefont
  {Rossell}, \citenamefont {Yang}, \citenamefont {You}, \citenamefont
  {Singh-Bhalla}, \citenamefont {Yang}, \citenamefont {He}, \citenamefont
  {Ramasse}, \citenamefont {Erni}, \citenamefont {Martin}, \citenamefont {Chu},
  \citenamefont {Pantelides}, \citenamefont {Pennycook},\ and\ \citenamefont
  {Ramesh}}]{yu_interface_2012}%
  \BibitemOpen
  \bibfield  {author} {\bibinfo {author} {\bibfnamefont {P.}~\bibnamefont
  {Yu}}, \bibinfo {author} {\bibfnamefont {W.}~\bibnamefont {Luo}}, \bibinfo
  {author} {\bibfnamefont {D.}~\bibnamefont {Yi}}, \bibinfo {author}
  {\bibfnamefont {J.~X.}\ \bibnamefont {Zhang}}, \bibinfo {author}
  {\bibfnamefont {M.~D.}\ \bibnamefont {Rossell}}, \bibinfo {author}
  {\bibfnamefont {C.-H.}\ \bibnamefont {Yang}}, \bibinfo {author}
  {\bibfnamefont {L.}~\bibnamefont {You}}, \bibinfo {author} {\bibfnamefont
  {G.}~\bibnamefont {Singh-Bhalla}}, \bibinfo {author} {\bibfnamefont {S.~Y.}\
  \bibnamefont {Yang}}, \bibinfo {author} {\bibfnamefont {Q.}~\bibnamefont
  {He}}, \bibinfo {author} {\bibfnamefont {Q.~M.}\ \bibnamefont {Ramasse}},
  \bibinfo {author} {\bibfnamefont {R.}~\bibnamefont {Erni}}, \bibinfo {author}
  {\bibfnamefont {L.~W.}\ \bibnamefont {Martin}}, \bibinfo {author}
  {\bibfnamefont {Y.~H.}\ \bibnamefont {Chu}}, \bibinfo {author} {\bibfnamefont
  {S.~T.}\ \bibnamefont {Pantelides}}, \bibinfo {author} {\bibfnamefont
  {S.~J.}\ \bibnamefont {Pennycook}}, \ and\ \bibinfo {author} {\bibfnamefont
  {R.}~\bibnamefont {Ramesh}},\ }\href {\doibase 10.1073/pnas.1117990109}
  {\bibfield  {journal} {\bibinfo  {journal} {Proceedings of the National
  Academy of Sciences}\ }\textbf {\bibinfo {volume} {109}},\ \bibinfo {pages}
  {9710} (\bibinfo {year} {2012})}\BibitemShut {NoStop}%
\bibitem [{\citenamefont {Takahashi}, \citenamefont {Kawasaki},\ and\
  \citenamefont {Tokura}(2001)}]{takahashi_interface_2001}%
  \BibitemOpen
  \bibfield  {author} {\bibinfo {author} {\bibfnamefont {K.~S.}\ \bibnamefont
  {Takahashi}}, \bibinfo {author} {\bibfnamefont {M.}~\bibnamefont {Kawasaki}},
  \ and\ \bibinfo {author} {\bibfnamefont {Y.}~\bibnamefont {Tokura}},\ }\href
  {\doibase 10.1063/1.1398331} {\bibfield  {journal} {\bibinfo  {journal}
  {Applied Physics Letters}\ }\textbf {\bibinfo {volume} {79}},\ \bibinfo
  {pages} {1324} (\bibinfo {year} {2001})}\BibitemShut {NoStop}%
\bibitem [{\citenamefont {Nichols}\ \emph {et~al.}(2016)\citenamefont
  {Nichols}, \citenamefont {Gao}, \citenamefont {Lee}, \citenamefont {Meyer},
  \citenamefont {Freeland}, \citenamefont {Lauter}, \citenamefont {Yi},
  \citenamefont {Liu}, \citenamefont {Haskel}, \citenamefont {Petrie},
  \citenamefont {Guo}, \citenamefont {Herklotz}, \citenamefont {Lee},
  \citenamefont {Ward}, \citenamefont {Eres}, \citenamefont {Fitzsimmons},\
  and\ \citenamefont {Lee}}]{nichols_emerging_2016}%
  \BibitemOpen
  \bibfield  {author} {\bibinfo {author} {\bibfnamefont {J.}~\bibnamefont
  {Nichols}}, \bibinfo {author} {\bibfnamefont {X.}~\bibnamefont {Gao}},
  \bibinfo {author} {\bibfnamefont {S.}~\bibnamefont {Lee}}, \bibinfo {author}
  {\bibfnamefont {T.~L.}\ \bibnamefont {Meyer}}, \bibinfo {author}
  {\bibfnamefont {J.~W.}\ \bibnamefont {Freeland}}, \bibinfo {author}
  {\bibfnamefont {V.}~\bibnamefont {Lauter}}, \bibinfo {author} {\bibfnamefont
  {D.}~\bibnamefont {Yi}}, \bibinfo {author} {\bibfnamefont {J.}~\bibnamefont
  {Liu}}, \bibinfo {author} {\bibfnamefont {D.}~\bibnamefont {Haskel}},
  \bibinfo {author} {\bibfnamefont {J.~R.}\ \bibnamefont {Petrie}}, \bibinfo
  {author} {\bibfnamefont {E.-J.}\ \bibnamefont {Guo}}, \bibinfo {author}
  {\bibfnamefont {A.}~\bibnamefont {Herklotz}}, \bibinfo {author}
  {\bibfnamefont {D.}~\bibnamefont {Lee}}, \bibinfo {author} {\bibfnamefont
  {T.~Z.}\ \bibnamefont {Ward}}, \bibinfo {author} {\bibfnamefont
  {G.}~\bibnamefont {Eres}}, \bibinfo {author} {\bibfnamefont {M.~R.}\
  \bibnamefont {Fitzsimmons}}, \ and\ \bibinfo {author} {\bibfnamefont {H.~N.}\
  \bibnamefont {Lee}},\ }\href {\doibase 10.1038/ncomms12721} {\bibfield
  {journal} {\bibinfo  {journal} {Nature Communications}\ }\textbf {\bibinfo
  {volume} {7}},\ \bibinfo {pages} {12721} (\bibinfo {year}
  {2016})}\BibitemShut {NoStop}%
\bibitem [{\citenamefont {Grutter}\ \emph {et~al.}(2015)\citenamefont
  {Grutter}, \citenamefont {Kirby}, \citenamefont {Gray}, \citenamefont
  {Flint}, \citenamefont {Alaan}, \citenamefont {Suzuki},\ and\ \citenamefont
  {Borchers}}]{grutter_electric_2015}%
  \BibitemOpen
  \bibfield  {author} {\bibinfo {author} {\bibfnamefont {A.}~\bibnamefont
  {Grutter}}, \bibinfo {author} {\bibfnamefont {B.}~\bibnamefont {Kirby}},
  \bibinfo {author} {\bibfnamefont {M.}~\bibnamefont {Gray}}, \bibinfo {author}
  {\bibfnamefont {C.}~\bibnamefont {Flint}}, \bibinfo {author} {\bibfnamefont
  {U.}~\bibnamefont {Alaan}}, \bibinfo {author} {\bibfnamefont
  {Y.}~\bibnamefont {Suzuki}}, \ and\ \bibinfo {author} {\bibfnamefont
  {J.}~\bibnamefont {Borchers}},\ }\href {\doibase
  10.1103/PhysRevLett.115.047601} {\bibfield  {journal} {\bibinfo  {journal}
  {Physical Review Letters}\ }\textbf {\bibinfo {volume} {115}},\ \bibinfo
  {pages} {047601} (\bibinfo {year} {2015})}\BibitemShut {NoStop}%
\bibitem [{\citenamefont {Freeland}\ \emph {et~al.}(2010)\citenamefont
  {Freeland}, \citenamefont {Chakhalian}, \citenamefont {Boris}, \citenamefont
  {Tonnerre}, \citenamefont {Kavich}, \citenamefont {Yordanov}, \citenamefont
  {Grenier}, \citenamefont {Zschack}, \citenamefont {Karapetrova},
  \citenamefont {Popovich}, \citenamefont {Lee},\ and\ \citenamefont
  {Keimer}}]{freeland_charge_2010}%
  \BibitemOpen
  \bibfield  {author} {\bibinfo {author} {\bibfnamefont {J.~W.}\ \bibnamefont
  {Freeland}}, \bibinfo {author} {\bibfnamefont {J.}~\bibnamefont
  {Chakhalian}}, \bibinfo {author} {\bibfnamefont {A.~V.}\ \bibnamefont
  {Boris}}, \bibinfo {author} {\bibfnamefont {J.-M.}\ \bibnamefont {Tonnerre}},
  \bibinfo {author} {\bibfnamefont {J.~J.}\ \bibnamefont {Kavich}}, \bibinfo
  {author} {\bibfnamefont {P.}~\bibnamefont {Yordanov}}, \bibinfo {author}
  {\bibfnamefont {S.}~\bibnamefont {Grenier}}, \bibinfo {author} {\bibfnamefont
  {P.}~\bibnamefont {Zschack}}, \bibinfo {author} {\bibfnamefont
  {E.}~\bibnamefont {Karapetrova}}, \bibinfo {author} {\bibfnamefont
  {P.}~\bibnamefont {Popovich}}, \bibinfo {author} {\bibfnamefont {H.~N.}\
  \bibnamefont {Lee}}, \ and\ \bibinfo {author} {\bibfnamefont
  {B.}~\bibnamefont {Keimer}},\ }\href {\doibase 10.1103/PhysRevB.81.094414}
  {\bibfield  {journal} {\bibinfo  {journal} {Physical Review B}\ }\textbf
  {\bibinfo {volume} {81}},\ \bibinfo {pages} {094414} (\bibinfo {year}
  {2010})}\BibitemShut {NoStop}%
\bibitem [{\citenamefont {Nanda}, \citenamefont {Satpathy},\ and\ \citenamefont
  {Springborg}(2007)}]{nanda_electron_2007}%
  \BibitemOpen
  \bibfield  {author} {\bibinfo {author} {\bibfnamefont {B.~R.~K.}\
  \bibnamefont {Nanda}}, \bibinfo {author} {\bibfnamefont {S.}~\bibnamefont
  {Satpathy}}, \ and\ \bibinfo {author} {\bibfnamefont {M.~S.}\ \bibnamefont
  {Springborg}},\ }\href {\doibase 10.1103/PhysRevLett.98.216804} {\bibfield
  {journal} {\bibinfo  {journal} {Physical Review Letters}\ }\textbf {\bibinfo
  {volume} {98}},\ \bibinfo {pages} {216804} (\bibinfo {year}
  {2007})}\BibitemShut {NoStop}%
\bibitem [{\citenamefont {Bhowal}\ and\ \citenamefont
  {Satpathy}(2019{\natexlab{a}})}]{bhowal_electronic_2019}%
  \BibitemOpen
  \bibfield  {author} {\bibinfo {author} {\bibfnamefont {S.}~\bibnamefont
  {Bhowal}}\ and\ \bibinfo {author} {\bibfnamefont {S.}~\bibnamefont
  {Satpathy}},\ }\href {\doibase 10.1103/PhysRevB.99.245145} {\bibfield
  {journal} {\bibinfo  {journal} {Physical Review B}\ }\textbf {\bibinfo
  {volume} {99}},\ \bibinfo {pages} {245145} (\bibinfo {year}
  {2019}{\natexlab{a}})}\BibitemShut {NoStop}%
\bibitem [{\citenamefont {Bhowal}\ and\ \citenamefont
  {Satpathy}(2019{\natexlab{b}})}]{bhowal_electric_2019}%
  \BibitemOpen
  \bibfield  {author} {\bibinfo {author} {\bibfnamefont {S.}~\bibnamefont
  {Bhowal}}\ and\ \bibinfo {author} {\bibfnamefont {S.}~\bibnamefont
  {Satpathy}},\ }\href {\doibase 10.1038/s41524-019-0198-8} {\bibfield
  {journal} {\bibinfo  {journal} {npj Computational Materials}\ }\textbf
  {\bibinfo {volume} {5}},\ \bibinfo {pages} {61} (\bibinfo {year}
  {2019}{\natexlab{b}})}\BibitemShut {NoStop}%
\bibitem [{\citenamefont {Lim}\ \emph {et~al.}(2020)\citenamefont {Lim},
  \citenamefont {Li}, \citenamefont {Huang}, \citenamefont {Chi}, \citenamefont
  {Zhou}, \citenamefont {Zeng}, \citenamefont {Omar}, \citenamefont {Feng},
  \citenamefont {Rusydi}, \citenamefont {Pennycook}, \citenamefont
  {Venkatesan},\ and\ \citenamefont {Ariando}}]{lim_emergent_2020}%
  \BibitemOpen
  \bibfield  {author} {\bibinfo {author} {\bibfnamefont {Z.~S.}\ \bibnamefont
  {Lim}}, \bibinfo {author} {\bibfnamefont {C.}~\bibnamefont {Li}}, \bibinfo
  {author} {\bibfnamefont {Z.}~\bibnamefont {Huang}}, \bibinfo {author}
  {\bibfnamefont {X.}~\bibnamefont {Chi}}, \bibinfo {author} {\bibfnamefont
  {J.}~\bibnamefont {Zhou}}, \bibinfo {author} {\bibfnamefont {S.}~\bibnamefont
  {Zeng}}, \bibinfo {author} {\bibfnamefont {G.~J.}\ \bibnamefont {Omar}},
  \bibinfo {author} {\bibfnamefont {Y.~P.}\ \bibnamefont {Feng}}, \bibinfo
  {author} {\bibfnamefont {A.}~\bibnamefont {Rusydi}}, \bibinfo {author}
  {\bibfnamefont {S.~J.}\ \bibnamefont {Pennycook}}, \bibinfo {author}
  {\bibfnamefont {T.}~\bibnamefont {Venkatesan}}, \ and\ \bibinfo {author}
  {\bibfnamefont {A.}~\bibnamefont {Ariando}},\ }\href {\doibase
  10.1002/smll.202004683} {\bibfield  {journal} {\bibinfo  {journal} {Small}\
  }\textbf {\bibinfo {volume} {16}},\ \bibinfo {pages} {2004683} (\bibinfo
  {year} {2020})}\BibitemShut {NoStop}%
\bibitem [{\citenamefont {Fujioka}\ \emph {et~al.}(2019)\citenamefont
  {Fujioka}, \citenamefont {Yamada}, \citenamefont {Kawamura}, \citenamefont
  {Sakai}, \citenamefont {Hirayama}, \citenamefont {Arita}, \citenamefont
  {Okawa}, \citenamefont {Hashizume}, \citenamefont {Hoshino},\ and\
  \citenamefont {Tokura}}]{fujioka_strong-correlation_2019}%
  \BibitemOpen
  \bibfield  {author} {\bibinfo {author} {\bibfnamefont {J.}~\bibnamefont
  {Fujioka}}, \bibinfo {author} {\bibfnamefont {R.}~\bibnamefont {Yamada}},
  \bibinfo {author} {\bibfnamefont {M.}~\bibnamefont {Kawamura}}, \bibinfo
  {author} {\bibfnamefont {S.}~\bibnamefont {Sakai}}, \bibinfo {author}
  {\bibfnamefont {M.}~\bibnamefont {Hirayama}}, \bibinfo {author}
  {\bibfnamefont {R.}~\bibnamefont {Arita}}, \bibinfo {author} {\bibfnamefont
  {T.}~\bibnamefont {Okawa}}, \bibinfo {author} {\bibfnamefont
  {D.}~\bibnamefont {Hashizume}}, \bibinfo {author} {\bibfnamefont
  {M.}~\bibnamefont {Hoshino}}, \ and\ \bibinfo {author} {\bibfnamefont
  {Y.}~\bibnamefont {Tokura}},\ }\href {\doibase 10.1038/s41467-018-08149-y}
  {\bibfield  {journal} {\bibinfo  {journal} {Nature Communications}\ }\textbf
  {\bibinfo {volume} {10}},\ \bibinfo {pages} {362} (\bibinfo {year}
  {2019})}\BibitemShut {NoStop}%
\bibitem [{\citenamefont {Masuko}\ \emph {et~al.}(2019)\citenamefont {Masuko},
  \citenamefont {Fujioka}, \citenamefont {Nakamura}, \citenamefont {Kawasaki},\
  and\ \citenamefont {Tokura}}]{masuko_strain-engineering_2019}%
  \BibitemOpen
  \bibfield  {author} {\bibinfo {author} {\bibfnamefont {M.}~\bibnamefont
  {Masuko}}, \bibinfo {author} {\bibfnamefont {J.}~\bibnamefont {Fujioka}},
  \bibinfo {author} {\bibfnamefont {M.}~\bibnamefont {Nakamura}}, \bibinfo
  {author} {\bibfnamefont {M.}~\bibnamefont {Kawasaki}}, \ and\ \bibinfo
  {author} {\bibfnamefont {Y.}~\bibnamefont {Tokura}},\ }\href {\doibase
  10.1063/1.5109582} {\bibfield  {journal} {\bibinfo  {journal} {APL
  Materials}\ }\textbf {\bibinfo {volume} {7}},\ \bibinfo {pages} {081115}
  (\bibinfo {year} {2019})}\BibitemShut {NoStop}%
\bibitem [{\citenamefont {Nishino}\ \emph
  {et~al.}(2018{\natexlab{a}})\citenamefont {Nishino}, \citenamefont {Kozuka},
  \citenamefont {Uchida}, \citenamefont {Kagawa},\ and\ \citenamefont
  {Kawasaki}}]{nishino_electrical_2018}%
  \BibitemOpen
  \bibfield  {author} {\bibinfo {author} {\bibfnamefont {R.}~\bibnamefont
  {Nishino}}, \bibinfo {author} {\bibfnamefont {Y.}~\bibnamefont {Kozuka}},
  \bibinfo {author} {\bibfnamefont {M.}~\bibnamefont {Uchida}}, \bibinfo
  {author} {\bibfnamefont {F.}~\bibnamefont {Kagawa}}, \ and\ \bibinfo {author}
  {\bibfnamefont {M.}~\bibnamefont {Kawasaki}},\ }\href {\doibase
  10.1063/1.5010391} {\bibfield  {journal} {\bibinfo  {journal} {Applied
  Physics Letters}\ }\textbf {\bibinfo {volume} {112}},\ \bibinfo {pages}
  {051602} (\bibinfo {year} {2018}{\natexlab{a}})}\BibitemShut {NoStop}%
\bibitem [{\citenamefont {Nishino}\ \emph
  {et~al.}(2018{\natexlab{b}})\citenamefont {Nishino}, \citenamefont {Kozuka},
  \citenamefont {Kagawa}, \citenamefont {Uchida},\ and\ \citenamefont
  {Kawasaki}}]{nishino_ferroelectric_2018}%
  \BibitemOpen
  \bibfield  {author} {\bibinfo {author} {\bibfnamefont {R.}~\bibnamefont
  {Nishino}}, \bibinfo {author} {\bibfnamefont {Y.}~\bibnamefont {Kozuka}},
  \bibinfo {author} {\bibfnamefont {F.}~\bibnamefont {Kagawa}}, \bibinfo
  {author} {\bibfnamefont {M.}~\bibnamefont {Uchida}}, \ and\ \bibinfo {author}
  {\bibfnamefont {M.}~\bibnamefont {Kawasaki}},\ }\href {\doibase
  10.1063/1.5047558} {\bibfield  {journal} {\bibinfo  {journal} {Applied
  Physics Letters}\ }\textbf {\bibinfo {volume} {113}},\ \bibinfo {pages}
  {143501} (\bibinfo {year} {2018}{\natexlab{b}})}\BibitemShut {NoStop}%
\bibitem [{\citenamefont {Nishino}\ \emph {et~al.}(2020)\citenamefont
  {Nishino}, \citenamefont {Fujita}, \citenamefont {Kagawa},\ and\
  \citenamefont {Kawasaki}}]{nishino_evolution_2020}%
  \BibitemOpen
  \bibfield  {author} {\bibinfo {author} {\bibfnamefont {R.}~\bibnamefont
  {Nishino}}, \bibinfo {author} {\bibfnamefont {T.~C.}\ \bibnamefont {Fujita}},
  \bibinfo {author} {\bibfnamefont {F.}~\bibnamefont {Kagawa}}, \ and\ \bibinfo
  {author} {\bibfnamefont {M.}~\bibnamefont {Kawasaki}},\ }\href {\doibase
  10.1038/s41598-020-67580-8} {\bibfield  {journal} {\bibinfo  {journal}
  {Scientific Reports}\ }\textbf {\bibinfo {volume} {10}},\ \bibinfo {pages}
  {10864} (\bibinfo {year} {2020})}\BibitemShut {NoStop}%
\bibitem [{\citenamefont {Ueno}\ \emph {et~al.}(2008)\citenamefont {Ueno},
  \citenamefont {Nakamura}, \citenamefont {Shimotani}, \citenamefont {Ohtomo},
  \citenamefont {Kimura}, \citenamefont {Nojima}, \citenamefont {Aoki},
  \citenamefont {Iwasa},\ and\ \citenamefont
  {Kawasaki}}]{ueno_electric-field-induced_2008}%
  \BibitemOpen
  \bibfield  {author} {\bibinfo {author} {\bibfnamefont {K.}~\bibnamefont
  {Ueno}}, \bibinfo {author} {\bibfnamefont {S.}~\bibnamefont {Nakamura}},
  \bibinfo {author} {\bibfnamefont {H.}~\bibnamefont {Shimotani}}, \bibinfo
  {author} {\bibfnamefont {A.}~\bibnamefont {Ohtomo}}, \bibinfo {author}
  {\bibfnamefont {N.}~\bibnamefont {Kimura}}, \bibinfo {author} {\bibfnamefont
  {T.}~\bibnamefont {Nojima}}, \bibinfo {author} {\bibfnamefont
  {H.}~\bibnamefont {Aoki}}, \bibinfo {author} {\bibfnamefont {Y.}~\bibnamefont
  {Iwasa}}, \ and\ \bibinfo {author} {\bibfnamefont {M.}~\bibnamefont
  {Kawasaki}},\ }\href {\doibase 10.1038/nmat2298} {\bibfield  {journal}
  {\bibinfo  {journal} {Nature Materials}\ }\textbf {\bibinfo {volume} {7}},\
  \bibinfo {pages} {855} (\bibinfo {year} {2008})}\BibitemShut {NoStop}%
\bibitem [{\citenamefont {Vistoli}\ \emph {et~al.}(2019)\citenamefont
  {Vistoli}, \citenamefont {Wang}, \citenamefont {Sander}, \citenamefont {Zhu},
  \citenamefont {Casals}, \citenamefont {Cichelero}, \citenamefont
  {Barth\'{e}l\'{e}my}, \citenamefont {Fusil}, \citenamefont {Herranz},
  \citenamefont {Valencia}, \citenamefont {Abrudan}, \citenamefont {Weschke},
  \citenamefont {Nakazawa}, \citenamefont {Kohno}, \citenamefont {Santamaria},
  \citenamefont {Wu}, \citenamefont {Garcia},\ and\ \citenamefont
  {Bibes}}]{vistoli_giant_2019}%
  \BibitemOpen
  \bibfield  {author} {\bibinfo {author} {\bibfnamefont {L.}~\bibnamefont
  {Vistoli}}, \bibinfo {author} {\bibfnamefont {W.}~\bibnamefont {Wang}},
  \bibinfo {author} {\bibfnamefont {A.}~\bibnamefont {Sander}}, \bibinfo
  {author} {\bibfnamefont {Q.}~\bibnamefont {Zhu}}, \bibinfo {author}
  {\bibfnamefont {B.}~\bibnamefont {Casals}}, \bibinfo {author} {\bibfnamefont
  {R.}~\bibnamefont {Cichelero}}, \bibinfo {author} {\bibfnamefont
  {A.}~\bibnamefont {Barth\'{e}l\'{e}my}}, \bibinfo {author} {\bibfnamefont
  {S.}~\bibnamefont {Fusil}}, \bibinfo {author} {\bibfnamefont
  {G.}~\bibnamefont {Herranz}}, \bibinfo {author} {\bibfnamefont
  {S.}~\bibnamefont {Valencia}}, \bibinfo {author} {\bibfnamefont
  {R.}~\bibnamefont {Abrudan}}, \bibinfo {author} {\bibfnamefont
  {E.}~\bibnamefont {Weschke}}, \bibinfo {author} {\bibfnamefont
  {K.}~\bibnamefont {Nakazawa}}, \bibinfo {author} {\bibfnamefont
  {H.}~\bibnamefont {Kohno}}, \bibinfo {author} {\bibfnamefont
  {J.}~\bibnamefont {Santamaria}}, \bibinfo {author} {\bibfnamefont
  {W.}~\bibnamefont {Wu}}, \bibinfo {author} {\bibfnamefont {V.}~\bibnamefont
  {Garcia}}, \ and\ \bibinfo {author} {\bibfnamefont {M.}~\bibnamefont
  {Bibes}},\ }\href {\doibase 10.1038/s41567-018-0307-5} {\bibfield  {journal}
  {\bibinfo  {journal} {Nature Physics}\ }\textbf {\bibinfo {volume} {15}},\
  \bibinfo {pages} {67} (\bibinfo {year} {2019})}\BibitemShut {NoStop}%
\bibitem [{\citenamefont {Xiang}\ \emph {et~al.}(2012)\citenamefont {Xiang},
  \citenamefont {Yamada}, \citenamefont {Akoh},\ and\ \citenamefont
  {Sawa}}]{xiang_phase_2012}%
  \BibitemOpen
  \bibfield  {author} {\bibinfo {author} {\bibfnamefont {P.-H.}\ \bibnamefont
  {Xiang}}, \bibinfo {author} {\bibfnamefont {H.}~\bibnamefont {Yamada}},
  \bibinfo {author} {\bibfnamefont {H.}~\bibnamefont {Akoh}}, \ and\ \bibinfo
  {author} {\bibfnamefont {A.}~\bibnamefont {Sawa}},\ }\href {\doibase
  10.1063/1.4768198} {\bibfield  {journal} {\bibinfo  {journal} {Journal of
  Applied Physics}\ }\textbf {\bibinfo {volume} {112}},\ \bibinfo {pages}
  {113703} (\bibinfo {year} {2012})}\BibitemShut {NoStop}%
\bibitem [{\citenamefont {Okamoto}\ \emph {et~al.}(2017)\citenamefont
  {Okamoto}, \citenamefont {Nichols}, \citenamefont {Sohn}, \citenamefont
  {Kim}, \citenamefont {Noh},\ and\ \citenamefont {Lee}}]{okamoto_charge_2017}%
  \BibitemOpen
  \bibfield  {author} {\bibinfo {author} {\bibfnamefont {S.}~\bibnamefont
  {Okamoto}}, \bibinfo {author} {\bibfnamefont {J.}~\bibnamefont {Nichols}},
  \bibinfo {author} {\bibfnamefont {C.}~\bibnamefont {Sohn}}, \bibinfo {author}
  {\bibfnamefont {S.~Y.}\ \bibnamefont {Kim}}, \bibinfo {author} {\bibfnamefont
  {T.~W.}\ \bibnamefont {Noh}}, \ and\ \bibinfo {author} {\bibfnamefont
  {H.~N.}\ \bibnamefont {Lee}},\ }\href {\doibase 10.1021/acs.nanolett.6b04107}
  {\bibfield  {journal} {\bibinfo  {journal} {Nano Letters}\ }\textbf {\bibinfo
  {volume} {17}},\ \bibinfo {pages} {2126} (\bibinfo {year}
  {2017})}\BibitemShut {NoStop}%
\bibitem [{\citenamefont {Hirai}\ \emph {et~al.}(2015)\citenamefont {Hirai},
  \citenamefont {Matsuno}, \citenamefont {Nishio-Hamane},\ and\ \citenamefont
  {Takagi}}]{hirai_semimetallic_2015}%
  \BibitemOpen
  \bibfield  {author} {\bibinfo {author} {\bibfnamefont {D.}~\bibnamefont
  {Hirai}}, \bibinfo {author} {\bibfnamefont {J.}~\bibnamefont {Matsuno}},
  \bibinfo {author} {\bibfnamefont {D.}~\bibnamefont {Nishio-Hamane}}, \ and\
  \bibinfo {author} {\bibfnamefont {H.}~\bibnamefont {Takagi}},\ }\href
  {\doibase 10.1063/1.4926723} {\bibfield  {journal} {\bibinfo  {journal}
  {Applied Physics Letters}\ }\textbf {\bibinfo {volume} {107}},\ \bibinfo
  {pages} {012104} (\bibinfo {year} {2015})}\BibitemShut {NoStop}%
\bibitem [{\citenamefont {Jackeli}\ and\ \citenamefont
  {Khaliullin}(2009)}]{jackeli_mott_2009}%
  \BibitemOpen
  \bibfield  {author} {\bibinfo {author} {\bibfnamefont {G.}~\bibnamefont
  {Jackeli}}\ and\ \bibinfo {author} {\bibfnamefont {G.}~\bibnamefont
  {Khaliullin}},\ }\href {\doibase 10.1103/PhysRevLett.102.017205} {\bibfield
  {journal} {\bibinfo  {journal} {Physical Review Letters}\ }\textbf {\bibinfo
  {volume} {102}},\ \bibinfo {pages} {017205} (\bibinfo {year}
  {2009})}\BibitemShut {NoStop}%
\bibitem [{\citenamefont {Onoda}, \citenamefont {Sugimoto},\ and\ \citenamefont
  {Nagaosa}(2008)}]{onoda_quantum_2008}%
  \BibitemOpen
  \bibfield  {author} {\bibinfo {author} {\bibfnamefont {S.}~\bibnamefont
  {Onoda}}, \bibinfo {author} {\bibfnamefont {N.}~\bibnamefont {Sugimoto}}, \
  and\ \bibinfo {author} {\bibfnamefont {N.}~\bibnamefont {Nagaosa}},\ }\href
  {\doibase 10.1103/PhysRevB.77.165103} {\bibfield  {journal} {\bibinfo
  {journal} {Physical Review B}\ }\textbf {\bibinfo {volume} {77}},\ \bibinfo
  {pages} {165103} (\bibinfo {year} {2008})}\BibitemShut {NoStop}%
\bibitem [{\citenamefont {Nagaosa}\ \emph {et~al.}(2010)\citenamefont
  {Nagaosa}, \citenamefont {Sinova}, \citenamefont {Onoda}, \citenamefont
  {MacDonald},\ and\ \citenamefont {Ong}}]{nagaosa_anomalous_2010}%
  \BibitemOpen
  \bibfield  {author} {\bibinfo {author} {\bibfnamefont {N.}~\bibnamefont
  {Nagaosa}}, \bibinfo {author} {\bibfnamefont {J.}~\bibnamefont {Sinova}},
  \bibinfo {author} {\bibfnamefont {S.}~\bibnamefont {Onoda}}, \bibinfo
  {author} {\bibfnamefont {A.~H.}\ \bibnamefont {MacDonald}}, \ and\ \bibinfo
  {author} {\bibfnamefont {N.~P.}\ \bibnamefont {Ong}},\ }\href {\doibase
  10.1103/RevModPhys.82.1539} {\bibfield  {journal} {\bibinfo  {journal}
  {Reviews of Modern Physics}\ }\textbf {\bibinfo {volume} {82}},\ \bibinfo
  {pages} {1539} (\bibinfo {year} {2010})}\BibitemShut {NoStop}%
\bibitem [{\citenamefont {Mohanta}\ \emph {et~al.}(2021)\citenamefont
  {Mohanta}, \citenamefont {Ok}, \citenamefont {Zhang}, \citenamefont {Miao},
  \citenamefont {Dagotto}, \citenamefont {Lee},\ and\ \citenamefont
  {Okamoto}}]{mohanta_semi-dirac_2021}%
  \BibitemOpen
  \bibfield  {author} {\bibinfo {author} {\bibfnamefont {N.}~\bibnamefont
  {Mohanta}}, \bibinfo {author} {\bibfnamefont {J.~M.}\ \bibnamefont {Ok}},
  \bibinfo {author} {\bibfnamefont {J.}~\bibnamefont {Zhang}}, \bibinfo
  {author} {\bibfnamefont {H.}~\bibnamefont {Miao}}, \bibinfo {author}
  {\bibfnamefont {E.}~\bibnamefont {Dagotto}}, \bibinfo {author} {\bibfnamefont
  {H.~N.}\ \bibnamefont {Lee}}, \ and\ \bibinfo {author} {\bibfnamefont
  {S.}~\bibnamefont {Okamoto}},\ }\href {\doibase 10.1103/PhysRevB.104.235121}
  {\bibfield  {journal} {\bibinfo  {journal} {Physical Review B}\ }\textbf
  {\bibinfo {volume} {104}},\ \bibinfo {pages} {235121} (\bibinfo {year}
  {2021})}\BibitemShut {NoStop}%
\bibitem [{\citenamefont {Ok}\ \emph {et~al.}(2021)\citenamefont {Ok},
  \citenamefont {Mohanta}, \citenamefont {Zhang}, \citenamefont {Yoon},
  \citenamefont {Okamoto}, \citenamefont {Choi}, \citenamefont {Zhou},
  \citenamefont {Briggeman}, \citenamefont {Irvin}, \citenamefont {Lupini},
  \citenamefont {Pai}, \citenamefont {Skoropata}, \citenamefont {Sohn},
  \citenamefont {Li}, \citenamefont {Miao}, \citenamefont {Lawrie},
  \citenamefont {Choi}, \citenamefont {Eres}, \citenamefont {Levy},\ and\
  \citenamefont {Lee}}]{ok_correlated_2021}%
  \BibitemOpen
  \bibfield  {author} {\bibinfo {author} {\bibfnamefont {J.~M.}\ \bibnamefont
  {Ok}}, \bibinfo {author} {\bibfnamefont {N.}~\bibnamefont {Mohanta}},
  \bibinfo {author} {\bibfnamefont {J.}~\bibnamefont {Zhang}}, \bibinfo
  {author} {\bibfnamefont {S.}~\bibnamefont {Yoon}}, \bibinfo {author}
  {\bibfnamefont {S.}~\bibnamefont {Okamoto}}, \bibinfo {author} {\bibfnamefont
  {E.~S.}\ \bibnamefont {Choi}}, \bibinfo {author} {\bibfnamefont
  {H.}~\bibnamefont {Zhou}}, \bibinfo {author} {\bibfnamefont {M.}~\bibnamefont
  {Briggeman}}, \bibinfo {author} {\bibfnamefont {P.}~\bibnamefont {Irvin}},
  \bibinfo {author} {\bibfnamefont {A.~R.}\ \bibnamefont {Lupini}}, \bibinfo
  {author} {\bibfnamefont {Y.-Y.}\ \bibnamefont {Pai}}, \bibinfo {author}
  {\bibfnamefont {E.}~\bibnamefont {Skoropata}}, \bibinfo {author}
  {\bibfnamefont {C.}~\bibnamefont {Sohn}}, \bibinfo {author} {\bibfnamefont
  {H.}~\bibnamefont {Li}}, \bibinfo {author} {\bibfnamefont {H.}~\bibnamefont
  {Miao}}, \bibinfo {author} {\bibfnamefont {B.}~\bibnamefont {Lawrie}},
  \bibinfo {author} {\bibfnamefont {W.~S.}\ \bibnamefont {Choi}}, \bibinfo
  {author} {\bibfnamefont {G.}~\bibnamefont {Eres}}, \bibinfo {author}
  {\bibfnamefont {J.}~\bibnamefont {Levy}}, \ and\ \bibinfo {author}
  {\bibfnamefont {H.~N.}\ \bibnamefont {Lee}},\ }\href {\doibase
  10.1126/sciadv.abf9631} {\bibfield  {journal} {\bibinfo  {journal} {Science
  Advances}\ }\textbf {\bibinfo {volume} {7}},\ \bibinfo {pages} {eabf9631}
  (\bibinfo {year} {2021})}\BibitemShut {NoStop}%
\end{thebibliography}%


\begin{thebibliography}{2}%
\makeatletter
\providecommand \@ifxundefined [1]{%
 \@ifx{#1\undefined}
}%
\providecommand \@ifnum [1]{%
 \ifnum #1\expandafter \@firstoftwo
 \else \expandafter \@secondoftwo
 \fi
}%
\providecommand \@ifx [1]{%
 \ifx #1\expandafter \@firstoftwo
 \else \expandafter \@secondoftwo
 \fi
}%
\providecommand \natexlab [1]{#1}%
\providecommand \enquote  [1]{``#1''}%
\providecommand \bibnamefont  [1]{#1}%
\providecommand \bibfnamefont [1]{#1}%
\providecommand \citenamefont [1]{#1}%
\providecommand \href@noop [0]{\@secondoftwo}%
\providecommand \href [0]{\begingroup \@sanitize@url \@href}%
\providecommand \@href[1]{\@@startlink{#1}\@@href}%
\providecommand \@@href[1]{\endgroup#1\@@endlink}%
\providecommand \@sanitize@url [0]{\catcode `\\12\catcode `\$12\catcode
  `\&12\catcode `\#12\catcode `\^12\catcode `\_12\catcode `\%12\relax}%
\providecommand \@@startlink[1]{}%
\providecommand \@@endlink[0]{}%
\providecommand \url  [0]{\begingroup\@sanitize@url \@url }%
\providecommand \@url [1]{\endgroup\@href {#1}{\urlprefix }}%
\providecommand \urlprefix  [0]{URL }%
\providecommand \Eprint [0]{\href }%
\providecommand \doibase [0]{http://dx.doi.org/}%
\providecommand \selectlanguage [0]{\@gobble}%
\providecommand \bibinfo  [0]{\@secondoftwo}%
\providecommand \bibfield  [0]{\@secondoftwo}%
\providecommand \translation [1]{[#1]}%
\providecommand \BibitemOpen [0]{}%
\providecommand \bibitemStop [0]{}%
\providecommand \bibitemNoStop [0]{.\EOS\space}%
\providecommand \EOS [0]{\spacefactor3000\relax}%
\providecommand \BibitemShut  [1]{\csname bibitem#1\endcsname}%
\let\auto@bib@innerbib\@empty
\bibitem [{\citenamefont {Groenendijk}\ \emph {et~al.}(2016)\citenamefont
  {Groenendijk}, \citenamefont {Manca}, \citenamefont {Mattoni}, \citenamefont
  {Kootstra}, \citenamefont {Gariglio}, \citenamefont {Huang}, \citenamefont
  {van Heumen},\ and\ \citenamefont {Caviglia}}]{groenendijk_epitaxial_2016}%
  \BibitemOpen
  \bibfield  {author} {\bibinfo {author} {\bibfnamefont {D.~J.}\ \bibnamefont
  {Groenendijk}}, \bibinfo {author} {\bibfnamefont {N.}~\bibnamefont {Manca}},
  \bibinfo {author} {\bibfnamefont {G.}~\bibnamefont {Mattoni}}, \bibinfo
  {author} {\bibfnamefont {L.}~\bibnamefont {Kootstra}}, \bibinfo {author}
  {\bibfnamefont {S.}~\bibnamefont {Gariglio}}, \bibinfo {author}
  {\bibfnamefont {Y.}~\bibnamefont {Huang}}, \bibinfo {author} {\bibfnamefont
  {E.}~\bibnamefont {van Heumen}}, \ and\ \bibinfo {author} {\bibfnamefont
  {A.~D.}\ \bibnamefont {Caviglia}},\ }\href {\doibase 10.1063/1.4960101}
  {\bibfield  {journal} {\bibinfo  {journal} {Applied Physics Letters}\
  }\textbf {\bibinfo {volume} {109}},\ \bibinfo {pages} {041906} (\bibinfo
  {year} {2016})}\BibitemShut {NoStop}%
\bibitem [{\citenamefont {Manca}\ \emph {et~al.}(2018)\citenamefont {Manca},
  \citenamefont {Groenendijk}, \citenamefont {Pallecchi}, \citenamefont
  {Autieri}, \citenamefont {Tang}, \citenamefont {Telesio}, \citenamefont
  {Mattoni}, \citenamefont {McCollam}, \citenamefont {Picozzi},\ and\
  \citenamefont {Caviglia}}]{manca_balanced_2018}%
  \BibitemOpen
  \bibfield  {author} {\bibinfo {author} {\bibfnamefont {N.}~\bibnamefont
  {Manca}}, \bibinfo {author} {\bibfnamefont {D.~J.}\ \bibnamefont
  {Groenendijk}}, \bibinfo {author} {\bibfnamefont {I.}~\bibnamefont
  {Pallecchi}}, \bibinfo {author} {\bibfnamefont {C.}~\bibnamefont {Autieri}},
  \bibinfo {author} {\bibfnamefont {L.~M.~K.}\ \bibnamefont {Tang}}, \bibinfo
  {author} {\bibfnamefont {F.}~\bibnamefont {Telesio}}, \bibinfo {author}
  {\bibfnamefont {G.}~\bibnamefont {Mattoni}}, \bibinfo {author} {\bibfnamefont
  {A.}~\bibnamefont {McCollam}}, \bibinfo {author} {\bibfnamefont
  {S.}~\bibnamefont {Picozzi}}, \ and\ \bibinfo {author} {\bibfnamefont
  {A.~D.}\ \bibnamefont {Caviglia}},\ }\href {\doibase
  10.1103/PhysRevB.97.081105} {\bibfield  {journal} {\bibinfo  {journal}
  {Physical Review B}\ }\textbf {\bibinfo {volume} {97}},\ \bibinfo {pages}
  {081105} (\bibinfo {year} {2018})}\BibitemShut {NoStop}%
\end{thebibliography}%
\newpage
\section*{Figures}
\textbf{}
\newline

\newpage
\textbf{}
\newline
\newline
\newline
\newline

\newpage
\textbf{}
\newline\newline
\newline
\newline
\newline

\begin{figure*}[h]
	\includegraphics[width=13cm]{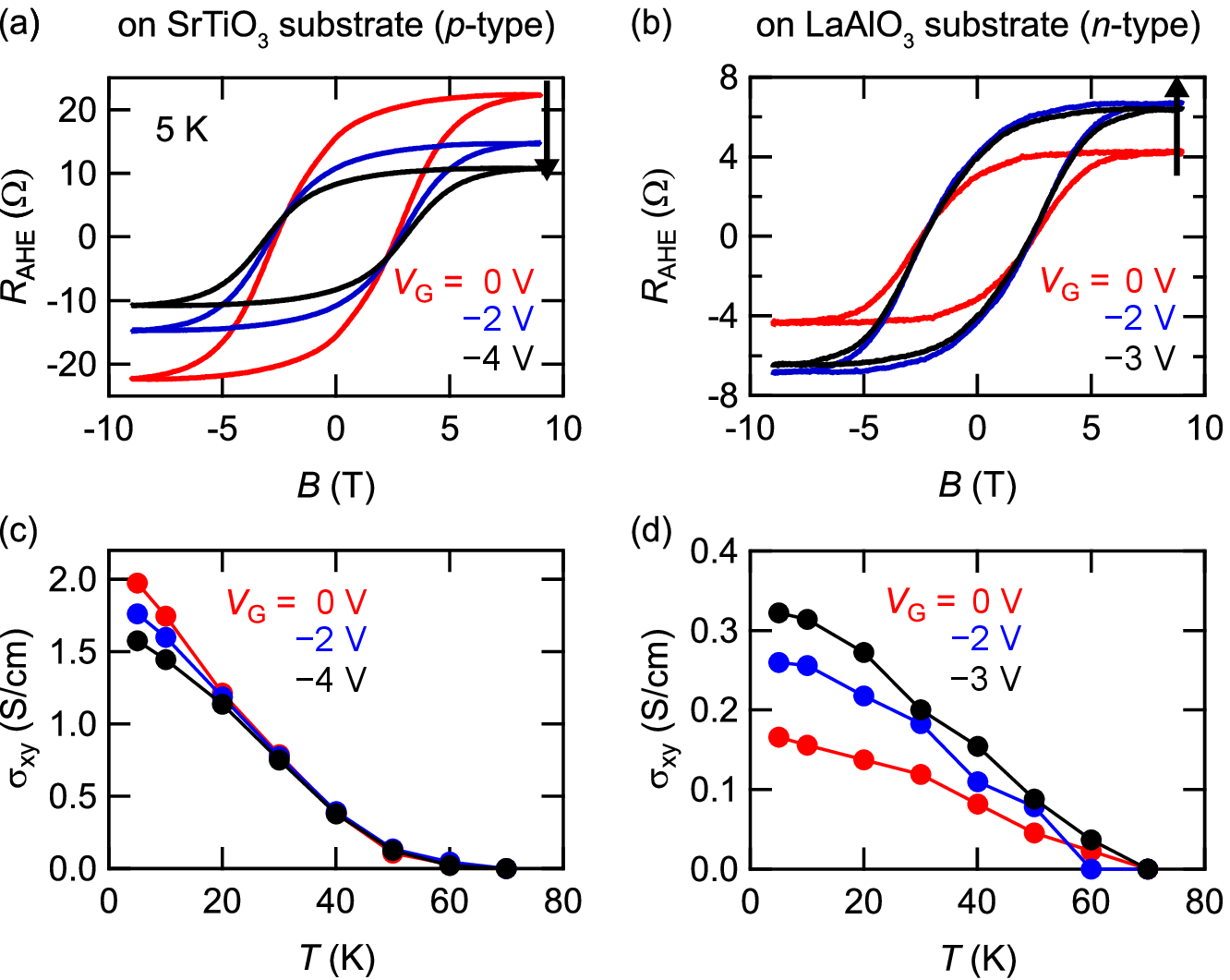}
	\colorcaption{\label{gate} 
	Magnetic field (B) dependence of anomalous Hall resistance ($R_\mathrm{AHE}$) at 5 K under several gate voltages ($V_\mathrm{G}$) for (a) \textit{p}- and (b) \textit{n}-type samples.
	Temperature dependence of anomalous Hall conductivity ($\sigma_\mathrm{xy}$) at 9 T under several gate voltages for (c) \textit{p}- and (d) \textit{n}-type samples.
			}
\end{figure*}

\newpage
\textbf{}
\newline\newline
\newline
\newline
\newline

\end{document}



\begin{center}
\textbf{\large Supplementary Materials:\\Electric field control of anomalous Hall effect in {\CIO}/{\CMO} heterostructure}
\end{center}
\renewcommand{\thepage}{\arabic{page}/{3}}
\setcounter{equation}{0}
\setcounter{figure}{0}
\setcounter{table}{0}
\setcounter{page}{1}
\makeatletter
\renewcommand{\theequation}{S\arabic{equation}}
\renewcommand{\thefigure}{S\arabic{figure}}
\renewcommand{\thetable}{S\arabic{table}}
\renewcommand{\bibnumfmt}[1]{[S#1] }
\renewcommand{\citenumfont}[1]{S#1}

\section{R\lowercase{eciprocal space mapping for} C\lowercase{a}I\lowercase{r}O$_3$/C\lowercase{a}M\lowercase{n}O$_3$ \lowercase{heterostructures grown on} S\lowercase{r}T\lowercase{i}O$_3$ \lowercase{and} L\lowercase{a}A\lowercase{l}O$_3$ \lowercase{substrates}}

\begin{figure*}[h]
	\includegraphics[width=11cm]{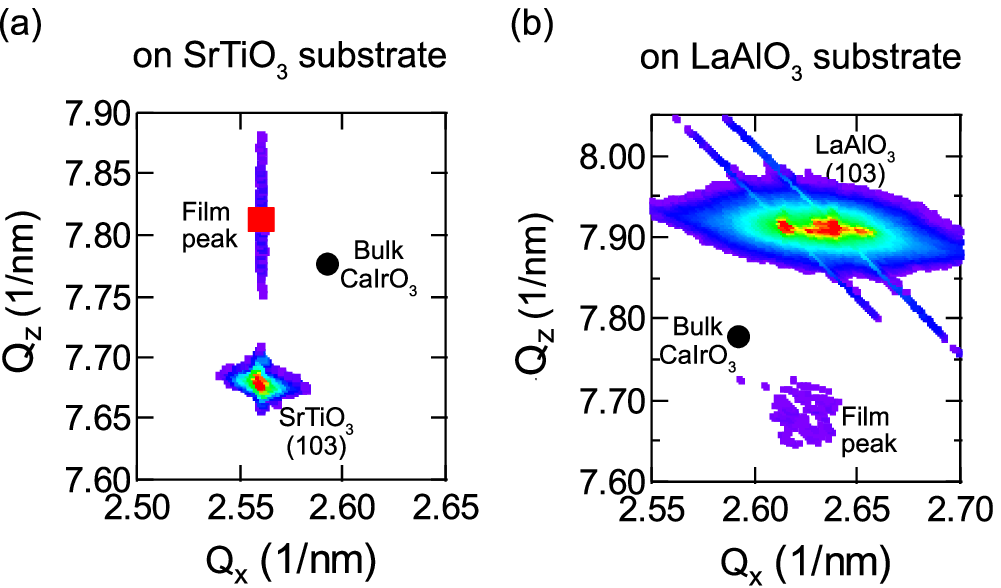}
	\colorcaption{\label{RSM}
		Reciprocal space mappings around (a) SrTiO$_3$ and (b) LaAlO$_3$ (103) peaks.
		Black circle denotes the peak position for the bulk CaIrO$_3$ calculated from pseudo-cubic lattice constant.
	}
\end{figure*}

Figures~\ref{RSM}(a) and ~\ref{RSM}(b) show the reciprocal space mapping images for CaIrO$_3$/CaMnO$_3$ heterostructures grown on SrTiO$_3$ (1.3\% tensile strain) and LaAlO$_3$ (1.7\% compressive strain) substrates, respectively.
CaIrO$_3$ films are fully strained to the substrates in both tensile and compressive strain cases.

\section{D\lowercase{evice structure for EDL gating and confirmation of no chemical reaction during device operation}}
\begin{figure*}[h]
	\includegraphics[width=11.5cm]{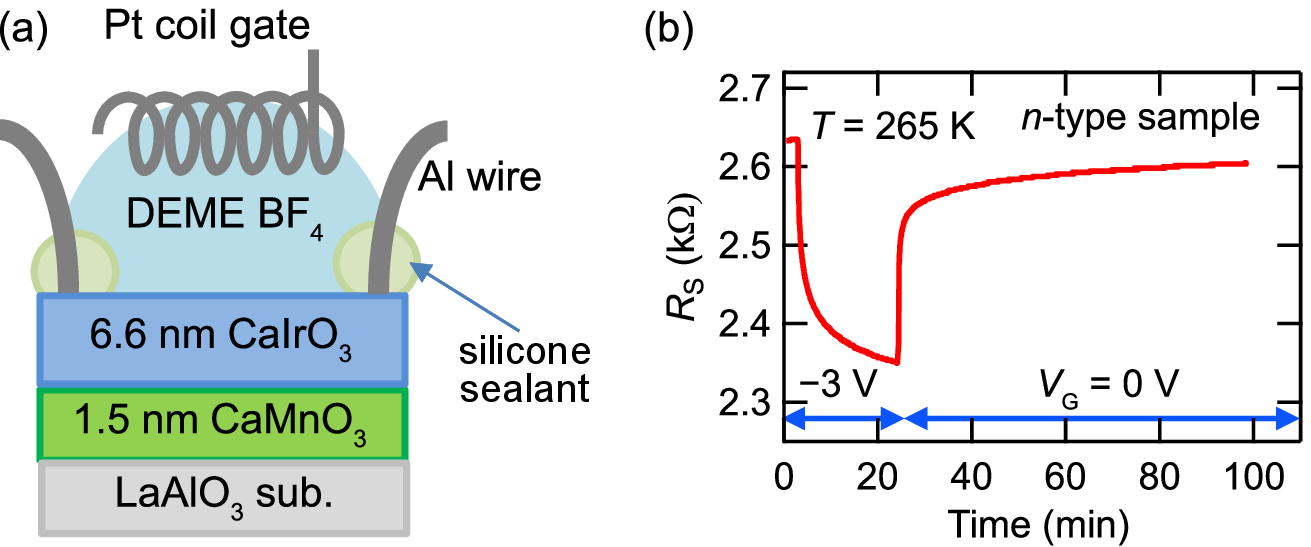}
	\colorcaption{\label{device}
		(a) Schematic diagram of the device structure for EDL gating. 
		(b) Time dependence of sheet resistance ($R_\mathrm{S}$) at 265 K for the heterostructure between gate voltage $V_\mathrm{G}$ = 0 V and $-3$ V.
	}
\end{figure*}

Figure~\ref{device}(a) illustrates a schematic diagram of the device structure for EDL gating.
As a gate dielectric, we employed an ionic liquid (IL), N, N-diethyl-N-(2-methoxyethyl)-N-methylammonium tetrafluoroborate (DEME-BF$_4$), in which a gate electrode of Pt coil is immersed.
For the electrical contacts, we utilized Al wire bonding.
Al wires are covered by silicone sealant to preserve a chemical reaction between the IL and Al wires.
For the device preparations, we did not use standard lithographic techniques because perovskite iridates are highly prone to be degraded in lithographic processes~\cite{groenendijk_epitaxial_2016}.
Figure~\ref{device}(b) shows an example of switching characteristics of the device between gate voltage $V_\mathrm{G}$ = 0 V and $-3$ V at 265 K. The reduced sheet resistance ($R_\mathrm{S}$) at $V_\mathrm{G}=-3$ V returns to the initial value after gating.
This reversible change indicates that EDL gating is an electrostatic modulation of the carrier density and does not cause any electrochemical reactions.

\section{T\lowercase{emperature dependence of sheet resistance for} C\lowercase{a}I\lowercase{r}O$_3$/C\lowercase{a}M\lowercase{n}O$_3$ \lowercase{heterostructures as a function of gate voltage}}
\begin{figure*}[h]
	\includegraphics[width=11.5cm]{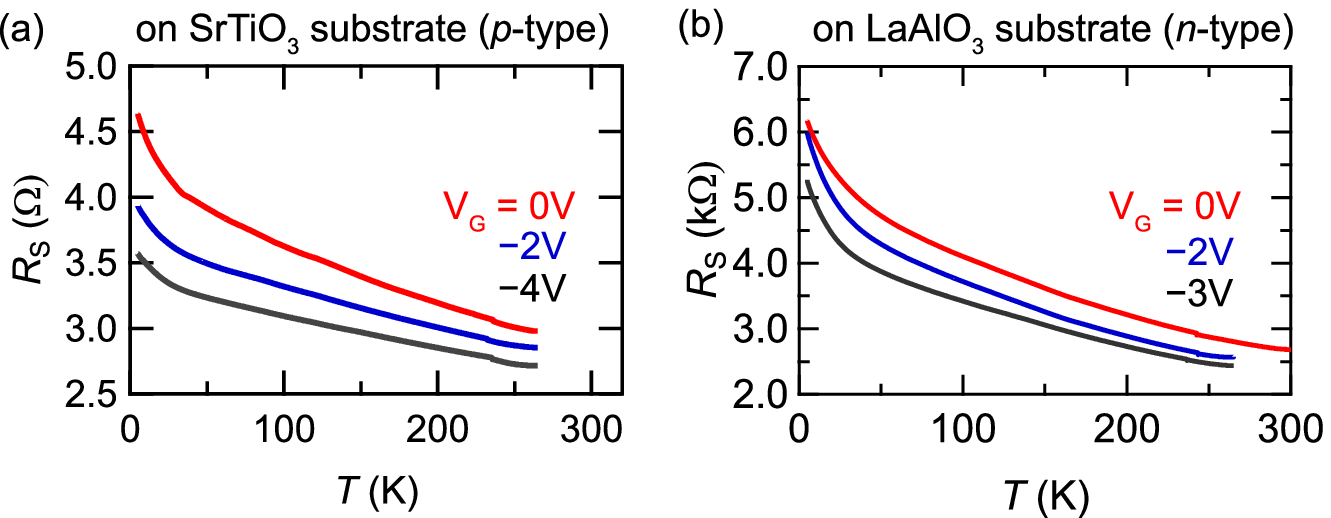}
	\colorcaption{\label{time}
		Temperature dependence of sheet resistance ($R_\mathrm{S}$) as a function of gate voltage $V_\mathrm{G}$ for (a) \textit{p}- and (b) \textit{n}-type samples.
	}
\end{figure*}

Figures~\ref{time}(a) and~\ref{time}(b) show temperature dependence of sheet resistance ($R_\mathrm{S}$) as a function of gate voltage $V_\mathrm{G}$ for \textit{p}- and \textit{n}-type heterostructures, respectively.
Negative $V_\mathrm{G}$ reduces the $R_\mathrm{S}$ in both carrier-type samples.
Since negative $V_\mathrm{G}$ is expected to deplete electron carrier, the behavior of the \textit{n}-type carrier sample seems to be opposite to expectation.
We speculate this complex dependence on $V_\mathrm{G}$ is related to semimetallic properties of perovskite iridates.
In fact, similar response to electric field is reported for SrIrO$_3$ previously~\cite{manca_balanced_2018}.

\textbf{}
\newline
\newline
\newline
\newline

\bibliography{CIO-CMO_NoNote}